\documentstyle[12pt]{article}

\begin{document}
\bibliographystyle{unsrt}

\begin{flushright} UMD-PP-94-117 \\

\today
\end{flushright}

\vspace{6mm}

\begin{center}

{\Large \bf Predictions for Proton Life-Time in Minimal Non-Supersymmetric
SO(10)
Models: {\it An Update}\footnote{Work supported by a grant from the National
Science
Foundation.}}\\ [6mm]
\vspace{10mm}

{\bf Dae-Gyu Lee and R. N. Mohapatra}\\
{\it Department of Physics, University of Maryland\\  College Park,
Maryland 20742}\\

and \\
{\bf M. K. Parida and Merostar Rani}\\
{\it Physics Department, North Eastern Hill University\\  P.O.Box 21,
Laitumkharh, Shillong-793003, India}\\ [4mm]

\vspace{10mm}

\end {center}

\begin{abstract}
We present our best estimates of the uncertainties due to
 heavy particle threshold corrections on the unification scale M$_{U}$,
 intermediate scale M$_{I}$, and coupling constant $\alpha_{U}$ in
 the minimal non-supersymmetric SO(10) models.  Using
 recent CERN e$^+$e$^-$ collider LEP data on sin$^2 \theta_{W}$ and
$\alpha_{strong}$ to obtain the  two-loop-level predictions for M$_{U}$ and
$\alpha_{U}$, we update the predictions for proton life-time in minimal
non-supersymmetric SO(10) models.

\end{abstract}

\newpage
\section{Introduction}
\hspace{8mm}
 The hypothesis of a single unified gauge symmetry describing
all forces and matter at very short distances is a very attractive one
from practical as well as aesthetic point of view. Right now
there are several good reasons to think that this gauge symmetry
may indeed be SO(10)\cite{rf1}. The most compelling argument
in favor of SO(10) comes from ways to explain\cite{rf2}
the observed deficit\cite{rf3} of the solar neutrino flux
compared to the predictions\cite{rf4} of the standard solar model
in terms of a two flavor MSW\cite{rf5} neutrino oscillation.
Consistent  understanding of  the data from
 all four experiments using the MSW oscillation hypothesis
requires the neutrino masses and mixings to lie in a very
narrow range of values. It was shown in a recent paper\cite{rf6},
that the minimal SO(10) theory that implements the see-saw
mechanism\cite{rf7} is a completely predictive theory in the neutrino sector
and predicts masses and mixings between $\nu_e$ and $\nu_{\mu}$
that are in this range.
In addition to this, there are many other highly desirable
 features of the SO(10) theory, such as fermion
unification into a single \{{\bf 16}\}-representation, a simple picture of
baryogenesis\cite{rf8}, asymptotic parity conservation of all
 interactions, etc. In view of these, we have undertaken a detailed
quantitative analysis of the symmetry breaking scales of this minimal model
in order to pinpoint its predictions for proton life time, especially
the unertainties in it arising from unknown Higgs masses in the theory.

Since  we are going to discuss the minimal SO(10) model, let us
explain what we mean by the word  `minimal'.
It stands for the fact that a) the Higgs sector is chosen to consist of the
smallest number of multiplets of SO(10) that is
 required for symmetry breaking and b) only those
fine-tunings of the parameters needed to achieve
 the desired gauge hierarchy are
imposed. The above fixes the Higgs mass spectrum of the model completely.

Before  we proceed further, we wish to make the following
important  remark about the minimal SO(10) model.
For a long time, it was thought that this model cannot be realistic since it
predicts\cite{rf9} the relations among fermion masses such as
 $m_{s}=m_{\mu}$ and $m_{d}=m_{e}$ at the GUT
 scale M$_{U}$, which after extrapolation to the weak scale, are in
complete disagreement with experiment.
However, it was shown in Ref.~\cite{rf6} that
 in the minimal SO(10) models where the small
neutrino masses arise from the see-saw mechanism\cite{rf7},
 there are additional
contributions to charged fermion masses, that solve this problem.
They arise from  the fact that the ({\bf 2,2,15})
submultiplet of the \{{\bf 126}\}-dim. Higgs
multiplet used in implementing the see-saw mechanism
 automatically acquires an induced VEV without additional fine-tuning.
These additional contributions correct the above mass relations in
such a way  as to restore agreement with observations.
The same theory, as mentioned above, also
 predicts the interesting values for neutrino masses and mixings
making the minimal SO(10) models not only completely realistic
but also testable by neutrino oscillation experiments to be
carried out soon.

Next, let us mention a word on our choice of
 non-supersymmetric version of the model.
While the question of gauge hierarchy certainly
prefers a supersymmetric (SUSY) SO(10) model,
in the absence of any evidence of supersymmetry at low energies
 as well as for the sake of simplicity alone, we believe that
 minimal non-SUSY SO(10) should be thoroughly explored and
confronted with experiments.

Another interesting point  that needs
 to be emphasized is that for the minimal set of Higgs multiplets, SO(10)
automatically breaks to the standard model via {\it only} one
 intermediate stage, that
consists of the left-right symmetric gauge group with or without the parity
symmetry\cite{rf10}, depending on the
 Higgs multiplet chosen to break SO(10). This leads
to the following four possibilities for the intermediate gauge symmetry: \\
\hspace*{10mm} A) G$_{224D}$ $\equiv$ SU(2)$_{L} \times$ SU(2)$_{R} \times$
SU(4)$_{C} \times$ D \\
\hspace*{10mm} B) G$_{224}$ $\equiv$  SU(2)$_{L} \times$ SU(2)$_{R} \times$
SU(4)$_{C}$ \\
\hspace*{10mm} C) G$_{2213D}$ $\equiv$  SU(2)$_{L} \times$ SU(2)$_{R} \times$
U(1)$_{B-L} \times$ SU(3)$_{C} \times$ D \\
\hspace*{10mm} D) G$_{2213}$ $\equiv$  SU(2)$_{L} \times$ SU(2)$_{R} \times$
U(1)$_{B-L} \times$ SU(3)$_{C}$ \\
Case A arises if the Higgs multiplet used to break is a single \{{\bf
54}\}-dimensional
one\cite{rf11}. Cases B and C arise if a single
 \{{\bf 210}\}-Higgs multiplet is used.
Depending on the range of the parameters in the Higgs
 potential, either case B or case C
arises as the intermediate symmetry\cite{rf12}. Case D arises when one uses a
combination of \{{\bf 45}\}- and \{{\bf 54}\}-dimensional Higgs
multiplets\cite{rf13}.
The rest of the symmetry breaking is implemented by a single \{{\bf
126}\}-dimensional
representation to break SU(2)$_{R} \times$ U(1)$_{B-L}$ as well as to
understand neutrino
masses and a single complex \{{\bf 10}\} to break the electroweak SU(2)$_L
\times$
U(1)$_{Y}$ down to U(1)$_{em}$.
These four cases therefore represent the four simplest and completely realistic
minimal
SO(10) models. In the rest of the paper, we present calculations
 of the predictions for proton life-time
($\tau_{p}$) in these models as well as the uncertainties in these predictions
due to unknown Higgs masses and the uncertainties in the low energy
input parameters,
 in order to see if  the next round of
proton decay search at Super-Kamiokande (SKAM)\cite{rf14} can test
this model.

\section{Computation of the Threshold Uncertainties in M$_{U}$ and M$_{I}$}
\hspace{8mm} The two main equations in our discussion are i) the two-loop
renormalization group equation for the evolution of the gauge coupling , i.e.,

\begin{eqnarray}
{\mbox{d} \alpha_{i} \over \mbox{d} t} =
 {\mbox{a}_{i} \over 2 \pi} \alpha_{i}^{2} + \sum_{j}
{\mbox{b}_{ij} \over 8 \pi^2} \alpha_{i}^{2} \alpha_{j},
\end{eqnarray}
and ii) the matching formula
 at the mass scale where the low energy symmetry group
enlarges\cite{rf15}

\begin{eqnarray}
{1 \over \alpha_{i}(M_{I})} = {1 \over \alpha_{I}(M_{I})}
-{\lambda_{i}^{I} \over 12 \pi}.
\end{eqnarray}

In Eqs. (1) and (2), $\alpha_{i}$ is the
 ``fine-structure" constant corresponding to the gauge
group G$_{i}$ and

\begin{eqnarray}
\lambda_{i}^{I}=\mbox{Tr } {\theta_{i}^{V}}^2 + \mbox{Tr } {\theta_{i}^{H}}^2
\mbox{ln}{\mbox{M}_{H} \over \mbox{M}_{I}},
\end{eqnarray}
where ${\theta_{i}^{H}}$ is the
 representation of the gauge group G$_{i}$ in the
representation of the Higgs submultiplet H.
 The expressions for a$_{i}$ and b$_{ij}$ for the
four cases are given in Table I\cite{rf16,rf17}.
 In deriving the values of a$_{i}$ and b$_{ij}$
in various cases as well as to obtain the threshold
 corrections $\lambda_{i}$, we need to
know the order of magnitude of the mass of
 the various Higgs submultiplets in the models.
We obtain these by invoking the survival hypothesis for the Higgs multiplets as
dictated by
the minimal fine tuning condition
 for gauge symmetry breaking\cite{rf18}. Using this
hypothesis, in Tables IIa-IId, we list the various
 Higgs multiplets whose masses are near the
relevant symmetry scales along with their contributions
to $\lambda_{i}^{I}$.

We proceed as follows:
 first using the
 two-loop equation, we derive the mean values for
the mass scales in various cases. These results already exist in the
literature\cite{rf16,rf17,rf19,rf20} based on the earlier LEP results.
 In Table III, we have presented their values from   Ref.~\cite{rf20},
which uses the inputs
$\alpha_1(M_Z)~=~0.016887\pm~0.000040$; $\alpha_2(M_Z)~=~0.03322\pm0.00025$;
$\alpha_3( M_Z)~=~0.120\pm0.007$, for further use in calcuating $\tau_{p}$.
These values of M$_{U}$ and M$_{I}$ were obtained
using analytic integration of Eq.~(1) which has been done exactly
for case A. For cases B, C, D, we have ignored terms whose effect in
the final result of the renormalization group equation is
 smaller than the error coming from low energy LEP data by a factor of
ten or more. We have also checked that inputting the most recent LEP
\cite{LEP} gives results for the mass scales
which are within the level of accuracy of our calculations.
For instance, for $M_U$ the changes are $10^{.08},10^{.01},10^{.06}$
, ~and~$ 10^{.14}$  for cases A, B, C and D respectively.
Then, we estimate the uncertainties in M$_{I}$ and M$_{U}$
 due to both the experimental uncertainties
 in the low energy parameters $\alpha_{s}$,
 sin$^2 \theta_{W}$ and $\alpha_{em}$
as well as the unknown Higgs boson masses. For the cases B and D, these were
discussed in Ref.~\cite{rf21} although we refine these
 uncertainties somewhat, but the
threshold uncertainties for cases A and C are new.
 These uncertainties are presented in
Table IV. We have allowed the Higgs masses
 to be between {\bf 1/10 to 10} times the
scale of the relevant symmetry breaking .

The formulas for the threshold effect
 on the mass scales M$_{I}$ and M$_{U}$, which were
used to obtain Table IV, are given below for
each symmetry breaking chain. We have
defined $\eta_{i}$ = ln (M$_{H_{i}}$/M), where
 M is the relevant gauge symmetry breaking scale
near M$_{H_{i}}$.

\noindent \underline{Model A}

\begin{eqnarray}
\Delta \mbox{ln}(M_{C_{+}} / M_{Z}) &=& 0 \eta_{10} +0 \eta_{126} +0 \eta_{54}
\nonumber \\
& & +0 \eta_{\phi} +0.431818 \eta_{R} -0.454545 \eta_{L} \nonumber   \\
\Delta \mbox{ln}(M_{U} / M_{Z}) &=& -0.04 \eta_{10} -0.08 \eta_{126}
  -0.04 \eta_{54}
\nonumber \\
& & +0.02 \eta_{\phi} -0.106818 \eta_{R} +0.194545 \eta_{L} \nonumber
\hspace{36mm} \mbox{(4A)}
\end{eqnarray}
\underline{Model B}

\begin{eqnarray}
\Delta \mbox{ln}(M_{C} / M_{Z}) &=&+ 0.0518135 \eta_{10} + 0.103627 \eta_{126}
+
0.0518135 \eta_{210} \nonumber \\
& & - 0.0259067 \eta_{\phi} +  1.12953 \eta_{R} - 1.29534 \eta_{\Delta_{L}}
\nonumber   \\
\Delta \mbox{ln}(M_{U} / M_{Z}) &=&- 0.0621762 \eta_{10} - 0.124352 \eta_{126}
- 0.0621762 \eta_{210} \nonumber \\
& & + 0.0310881 \eta_{\phi} - 0.40544   \eta_{R} + 0.554404 \eta_{\Delta_{L}}
\nonumber
\hspace{28mm} \mbox{(4B)}
\end{eqnarray}
\underline{Model C}

\begin{eqnarray}
\Delta \mbox{ln}(M_{R_{+}} / M_{Z}) &=& +0.0952381 \eta_{10} -0.0952381
\eta_{126}
+0\eta_{210} \nonumber \\
& &  -0.047619 \eta_{\phi} +0.190476 \eta_{R1} -0.142857 \eta_{L1} \nonumber
\\
\Delta \mbox{ln}(M_{U} / M_{Z}) &=& -0.0544218 \eta_{10} -0.159864 \eta_{126}
-0.0357143 \eta_{210} \nonumber \\
& & +0.0272109 \eta_{\phi} +0.0340136  \eta_{R1} +0.117347 \eta_{L1} \nonumber
\hspace{20mm} \mbox{(4C)}
\end{eqnarray}
\underline{Model D}

\begin{eqnarray}
\Delta \mbox{ln}(M_{R} / M_{Z}) &=& +0.124138 \eta_{10} -0.0827586 \eta_{126}
+0.00689655 \eta_{45} \nonumber \\
& & -0.062069 \eta_{\phi} +0.22069 \eta_{R1}  -0.193103 \eta_{H_{\Delta}}
\nonumber  \\
\Delta \mbox{ln}(M_{U} / M_{Z}) &=& -0.0781609 \eta_{10} -0.170115 \eta_{126}
-0.0413793 \eta_{45} \nonumber \\
& &+0.0390805  \eta_{\phi} +0.0091954 \eta_{R1} +0.158621 \eta_{H_{\Delta}}
\nonumber
\hspace{22mm} \mbox{(4D)}
\end{eqnarray}

In obtaining the above equations, we have assumed that the
particles from a single SO(10) representation which have masses
of the same order are degenerate. This is the same assumption
as in Ref.~\cite{rf21}.
Before proceeding to give our predictions for proton life-time, few comments
are in order. \\
a) We want to clarify how we get the uncertainties
 presented in Table IV. First, as already
mentioned, we chose M$_{H}$/M$_{I}$ or M$_{H}$/M$_{U}$ to
 vary between 10$^{-1}$
and 10$^{+1}$. The maximum values of the uncertainties
 are obtained by allowing the
different $\eta$'s to vary independently to their extreme
 values that lead to the largest
positive or negative uncertainty.
The only exception to this is the two parameters $\eta_{\Delta{_L}}$ and
$\eta_{H_{\Delta}}$, which are always kept negative. (See below.)
Secondly, in chains A and C we do not assume
 that the left- and right-handed Higgs
submultiplets to have same mass (that would lead to $\eta_{L}=\eta_{R}$).
The reason is that since the masses are close to the
 intermediate scale, where left-right
symmetry is broken, the multiplets need not necessarily be degenerate.
If we assumed the degeneracy, there would be
 cancellation between $\eta_{L}$ and
$\eta_{R}$ terms reducing the threshold uncertainties\cite{rf22};
 the uncertainties we
present in table IV are, therefore, most conservative. \\
b) In cases B and D, since D-parity is broken
 at the GUT scale, the masses of $\Delta_{L}$
in Table IIb-(1) and H$_{\Delta}$ in Table IId-(1) are
 always above the scale M$_{I}$, but
below M$_{U}$\cite{rf21}.
Although {\it a priori}  $M_{\Delta_{L}}$ could be bigger than M$_{U}$,
we have kept it smaller in presenting the uncertainty in $\tau_p$.
Therefore,   $\eta_{\Delta_{L}}$
 and $\eta_{H_{\Delta}}$ in Eqs.(4B) and (4D) are always negative,
since we use M = M$_{U}$ to define them. In any case, from an
experimental point of view, the upper value of the uncertainty
is not too relevant.\\
c) The first set of entries in Table IV is obtained
 by maximizing the uncertainty in M$_{I}$
whereas the second set is obtained by  reversing this procedure. \\
d) Note that, in case A, the intermediate mass scale
 M$_{I}$ and the unification scale
M$_{U}$ are so close that one might think of
 this as an almost single step breaking.
This is similar to the D-parity broken scenario (case B) recently discussed in
Ref.~\cite{rf23}.
For proton life-time estimate, this is inconsequential.

\section{Predictions for Proton Life-Time}
\hspace{8mm} Now, we present our predictions
 for proton life-time in the four SO(10)
models A - D. For this purpose, we need the values
 of M$_{U}$ and $\alpha_{U}$ and
remember that in SO(10) there are extra gauge bosons
 contributing to proton decay
compared to the SU(5) model. We use the following formula from the review by
Langacker\cite{rf24}, where the original literature can be found. We write
\begin{eqnarray}
\tau_{p} = \tau_{p}^{(0)} F_{p}, \nonumber
\end{eqnarray}
where $F_{p}$ denotes the uncertainty arising
 from threshold corrections as well as the
experimental errors in $\alpha_{s}$, $\alpha_{em}$, and $sin^2 \theta_{W}$.
{}From Ref.~\cite{rf24}, we get for $\tau_{p}^{(0)}$
\begin{eqnarray}
\hspace*{20mm} \tau_{p \rightarrow e^+\pi^0}^{(0)}=
{5 \over 8}\left({\alpha_{U}^{SU(5)} \over
\alpha_{U}^{SO(10)}} \right)^2 \times 4.5 \times 10^{29 \pm.7}
 \left({M_{U} \over 2.1 \times
10^{14} \mbox{GeV} }\right)^4 \mbox{yrs.} \hspace*{15mm} (5) \nonumber
\end{eqnarray}

Including the $F_{p}$-factors, we present below
 the predictions for proton life-time in
SO(10) (noting that $\alpha_{U}^{SU(5)} \approx \alpha_{U}^{SO(10)}$).
The first
uncertainty in the predictions below arises from the
 proton decay matrix element
evaluation whereas the second and the third ones come
from LEP data and threshold
correction, respectively\cite{rf26}.\\
\underline{Model A}
\begin{eqnarray}
\tau_{p \rightarrow e^+\pi^0}=1.44 \times
10^{32.1 \pm .7 \pm 1.0 \pm 1.9} \mbox{yrs.}
\nonumber
\end{eqnarray}
\underline{Model B}
\begin{eqnarray}
\tau_{p \rightarrow e^+\pi^0}=1.44 \times 10^{37.4 \pm .7 \pm 1.0
_{-5.0}^{+.5}} \mbox{yrs.}
\nonumber
\end{eqnarray}
\underline{Model C}
\begin{eqnarray}
\tau_{p \rightarrow e^+\pi^0}=1.44 \times 10^{34.2 \pm .7 \pm .8 \pm 1.7}
\mbox{yrs.}
\nonumber
\end{eqnarray}
\underline{Model D}
\begin{eqnarray}
\tau_{p \rightarrow e^+\pi^0}=1.44 \times 10^{37.7 \pm .7 \pm .9 _{-2.0}^{+.7}}
\mbox{yrs.}
\nonumber
\end{eqnarray}

\section{Conclusion}
\hspace*{8mm} In conclusion, we have computed the threshold
 uncertainties in both the intermediate and
the unification scales for all four possible minimal
 non-supersymmetric SO(10) models A -
D. We then update the predictions for proton life-time
 in all these cases including the most
conservative estimates for the threshold uncertainties in it.
We see that for case A,
$\tau_{p}$ is very much within the range of Super-Kamiokande
 search even without
threshold corrections.
On the other hand, for cases B and C, the threshold uncertainties
have the effect of bring it within the range of SKAM search.

\newpage
\section*{Table Caption}
\noindent Table I: One- and two-loop $\beta$-function coefficients for models A
- D.

\vspace*{3mm}

\noindent Table IIa: The heavy Higgs content of the model A.
 The G$_{224}$ submultiplets
in (1) acquire masses when SO(10) is broken, while the G$_{123}$
 submultiplets in (2)
become massive when G$_{224D}$ is broken.
The multiplets $\phi$, R$_{i}$, and L$_{i}$ in
(2) arise from $\phi$(2,2,0) in \{{\bf 10}\}, from $\Delta_{R}$(1,3,10), and
$\Delta_{L}$(3,1,$\overline{10}$) in
 \{{\bf 126}\}, respectively. Also listed, in the extreme
right column of the tables, are
 the threshold contributions $\lambda_{i}$ of the different multiplets.

\vspace*{3mm}

\noindent Table IIb: The heavy Higgs particles in the model B
 whose intermediate symmetry
is G$_{224}$. The particles whose masses are on the order
 of M$_{U}$ are listed in (1),
and the particles on the order of M$_{I}$ are listed in (2).
 Also listed are their threshold contributions.
\vspace*{3mm}

\noindent Table IIc: All the heavy Higgs particles in the model C
 whose intermediate symmetry
is G$_{2213D}$. The submultiplets with masses of order M$_{U}$
 are presented in (1), and
the particles with masses of order M$_{I}$ are listed in (2).
 The entries in the extreme right
column denote their threshold contributions.
\vspace*{3mm}

\noindent Table IId: All the heavy Higgs particles in the model D
 whose intermediate symmetry
is G$_{2213}$. The submultiplets with masses of
 order M$_{U}$ are presented in (1), and
the particles with masses of order M$_{I}$ are listed in (2).
 The entries in the far right column give their $\lambda_{i}$ contributions.
\vspace*{3mm}

\noindent Table III: The values of $M_U$, $M_I$ and $\alpha_U$ obtained
by solving the two-loop renormalization group equation Eq. (1) for
models A - D. The results were taken from Ref.~\cite{rf20}.
\vspace*{3mm}

\noindent Table IV: The threshold uncertainties due to
the difference between symmetry
breaking scale and masses of Higgs bosons on the order of that scale.
 For the cases where threshold effects in M$_{I}$ and M$_{U}$ are
 maximized, corresponding threshold uncertainties are given in the
 first two lines and the last two lines, respectively.

\newpage
\begin{center}
Table I. \\
\begin{tabular}{ ccc } \hline\hline
Model & a$_{i}$ & b$_{ij}$ \\ \hline
A &$\{ {{11}\over 3},{{11}\over 3},-{{14}\over 3}\}$
&$\{ \{ {{584}\over 3},3,{{765}\over 2}\} ,
  \{ 3,{{584}\over 3},{{765}\over 2}\} ,
  \{ {{153}\over 2},{{153}\over 2},{{1759}\over 6}\} \}$ \\

B &$\{ -3,{{11}\over 3},-{{23}\over 3}\}$
&$\{ \{ 8,3,{{45}\over 2}\} ,
  \{ 3,{{584}\over 3},{{765}\over 2}\} ,
  \{ {9\over 2},{{153}\over 2},{{643}\over 6}\} \}$ \\

C &$\{ -{7\over 3},-{7\over 3},7,-7\}$
&$\{ \{ {{80}\over 3},3,{{27}\over 2},12\} ,
  \{ 3,{{80}\over 3},{{27}\over 2},12\} ,
  \{ {{81}\over 2},{{81}\over 2},{{115}\over 2},4\} ,
  \{ {9\over 2},{9\over 2},{1\over 2},-26\} \}$ \\
D &$\{ -3,-{7\over 3},{{11}\over 2},-7\}$
&$\{ \{ 8,3,{3\over 2},12\} ,
  \{ 3,{{80}\over 3},{{27}\over 2},12\} ,
  \{ {9\over 2},{{81}\over 2},{{61}\over 2},4\} ,
  \{ {9\over 2},{9\over 2},{1\over 2},-26\} \}$ \\ \hline \hline
\end{tabular}

\end{center}

\begin{center}
Table IIa-(1). \\
\begin{tabular}{ ccc } \hline\hline
SO(10) representation & G$_{224}$ submultiplet &
\{$\lambda_{2L}^{U}$,$\lambda_{2R}^{U}$,$\lambda_{4C}^{U}$\} \\ \hline
{\bf 10} &H(1,1,6) &\{ 0, 0, 2\} \\
{\bf 126} & $\zeta_{0}$(2,2,15) &\{30,30,32\} \\
          &S(1,1,6) &\{ 0, 0, 2\} \\
{\bf 54} &S$_{\Sigma}$(3,3,1) &\{ 6, 6, 0\} \\
          &S$_{\zeta}$(1,1,20$^{'}$) &\{ 0, 0, 8\} \\
          &S$_{+}$(1,1,1) &\{ 0, 0, 0\} \\ \hline \hline
\end{tabular}

\end{center}
\newpage
\begin{center}
Table IIa-(2). \\
\begin{tabular}{ ccc } \hline\hline
SO(10) representation & G$_{123}$ submultiplet &
\{$\lambda_{1Y}^{I}$,$\lambda_{2L}^{I}$,$\lambda_{3C}^{I}$\} \\ \hline
{\bf 10}

&$\phi$($-{1 \over 2} \sqrt{3 \over 5}$,2,1) &\{${3 \over 5}$, 1, 0\} \\
{\bf 126}

&R$_{1}$($-{ 2} \sqrt{3 \over 5}$,1,1) &\{${24 \over 5}$, 0, 0\}  \\

&R$_{2}$($+{1 \over 3} \sqrt{3 \over 5}$,1,3) &\{${2 \over 5}$, 0, 1\} \\

&R$_{3}$($-{4 \over 3} \sqrt{3 \over 5}$,1,3) &\{${32 \over 5}$, 0, 1\} \\

&R$_{4}$($-{1 \over 3} \sqrt{3 \over 5}$,1,6) &\{${4 \over 5}$, 0, 5\} \\

&R$_{5}$($+{2 \over 3} \sqrt{3 \over 5}$,1,6) &\{${16 \over 5}$, 0, 5\} \\

&R$_{6}$($-{4 \over 3} \sqrt{3 \over 5}$,1,6) &\{${64 \over 5}$, 0, 5\} \\

&L$_{1}$($+ \sqrt{3 \over 5}$,3,1) &\{${18 \over 5}$, 4, 0\} \\

&L$_{2}$($+{1 \over 3} \sqrt{3 \over 5}$,3,$\overline{3}$) &\{${6 \over 5}$,12,
3\} \\

&L$_{3}$($-{1 \over 3} \sqrt{3 \over 5}$,3,$\overline{6}$) &\{${12 \over
5}$,24, 15\}   \\
\hline \hline
\end{tabular}

\end{center}

\begin{center}
Table IIb-(1). \\
\begin{tabular}{ ccc } \hline\hline
SO(10) representation & G$_{224}$ submultiplet &
\{$\lambda_{2L}^{U}$,$\lambda_{2R}^{U}$,$\lambda_{4C}^{U}$\} \\ \hline
{\bf 10} &H(1,1,6) &\{ 0, 0, 2\} \\
{\bf 126} & $\zeta_{0}$(2,2,15) &\{30,30,32\} \\
          &S(1,1,6) &\{ 0, 0, 2\} \\
          &$\Delta_{L}$(3,1,$\overline{10}$) &\{40, 0,18\} \\
{\bf 210} &$\Sigma_{L}$(3,1,15) &\{30, 0,12\} \\
          &$\Sigma_{R}$(1,3,15) &\{ 0,30,12\} \\
          &$\zeta_{1}$(2,2,10) &\{10,10,12\} \\
          &$\zeta_{2}$(2,2,$\overline{10}$) &\{10,10,12\} \\
          &$\zeta_{3}$(1,1,15) &\{ 0, 0, 4\} \\
          &S$^{'}$(1,1,1) &\{ 0, 0, 0\} \\ \hline \hline
\end{tabular}

\end{center}
\newpage
\begin{center}
Table IIb-(2). \\
\begin{tabular}{ ccc } \hline\hline
 G$_{123}$ submultiplet  \\ \hline
$\phi$, R$_{1}$, R$_{2}$, R$_{3}$, R$_{4}$, R$_{5}$, R$_{6}$ in TableIIa-(2) \\
\hline
\hline
\end{tabular}

\end{center}

\newpage
{\footnotesize
\begin{center}
Table IIc-(1). \\
\begin{tabular}{ ccc } \hline\hline
SO(10) representation & G$_{2213}$ submultiplet &
\{$\lambda_{2L}^{U}$,$\lambda_{2R}^{U}$,$\lambda_{1X}^{U}$,$\lambda_{3C}^{U}$\}
\\
\hline
{\bf 10}

&T$_{1}$(1,1,$+{1 \over 3} \sqrt{3 \over 2}$, $\overline{3}$) &\{ 0, 0, 1, 1\}
\\
&T$_{2}$(1,1,$-{1 \over 3} \sqrt{3 \over 2}$, 3) &\{ 0, 0, 1, 1\} \\
{\bf 126}

&H$_{1R}$(1,3,$+{1 \over 3} \sqrt{3 \over 2}$, 6) &\{ 0,24, 6,15\} \\
&H$_{1L}$(3,1,$-{1 \over 3} \sqrt{3 \over 2}$,$\overline{6}$) &\{24, 0, 6,15\}
\\
&H$_{2R}$(1,3,$-{1 \over 3} \sqrt{3 \over 2}$, 3) &\{ 0, 4, 3, 3\} \\
&H$_{2L}$(3,1,$+{1 \over 3} \sqrt{3 \over 2}$, $\overline{3}$) &\{ 4, 0, 3, 3\}
\\
&H$_{3}$(2,2,$+{2 \over 3} \sqrt{3 \over 2}$, 3) &\{ 6, 6,16, 4\} \\
&H$_{4}$(2,2,$-{2 \over 3} \sqrt{3 \over 2}$,$\overline{3}$) &\{ 6, 6,16, 4\}
\\
&H$_{5}$(2,2,0,8) &\{16,16, 0,24\} \\
&H$_{6}$(2,2,0,1) &\{ 2, 2, 0, 0\} \\
&H$_{7}$(1,1,$+{1 \over 3} \sqrt{3 \over 2}$, $\overline{3}$) &\{ 0, 0, 1, 1\}
\\
&H$_{8}$(1,1,$-{1 \over 3} \sqrt{3 \over 2}$, 3) &\{ 0, 0, 1, 1\} \\
{\bf 210}

&B$_{L1}$(3,1,0,8) &\{16, 0, 0, 9\} \\
&B$_{R1}$(1,3,0,8) &\{ 0,16, 0, 9\} \\
&B$_{L2}$(3,1,$-{2 \over 3} \sqrt{3 \over 2}$, $\overline{3}$) &\{ 6, 0, 6,${3
\over 2}$\} \\
&B$_{R2}$(1,3,$-{2 \over 3} \sqrt{3 \over 2}$, $\overline{3}$) &\{ 0, 6, 6,${3
\over 2}$\} \\
&B$_{L3}$(3,1,$+{2 \over 3} \sqrt{3 \over 2}$, 3) &\{ 6, 0, 6,${3 \over 2}$\}
\\
&B$_{R3}$(1,3,$+{2 \over 3} \sqrt{3 \over 2}$, 3) &\{ 0, 6, 6,${3 \over 2}$\}
\\
&B$_{L4}$(3,1,0,1) &\{ 2, 0, 0, 0\} \\
&B$_{R4}$(1,3,0,1) &\{ 0, 2, 0, 0\} \\
&B$_{5}$(2,2,$+{1 \over 3} \sqrt{3 \over 2}$,6) &\{ 6, 6, 4,10\} \\
&B$_{6}$(2,2,$-{1 \over 3} \sqrt{3 \over 2}$,$\overline{6}$) &\{ 6, 6, 4,10\}
\\
&B$_{7}$(2,2,$-{1 \over 3} \sqrt{3 \over 2}$,3) &\{ 3, 3, 2, 2\} \\
&B$_{8}$(2,2,$+{1 \over 3} \sqrt{3 \over 2}$,$\overline{3}$) &\{ 3, 3, 2, 2\}
\\
&B$_{9}$(2,2,$- \sqrt{3 \over 2}$, 1) &\{ 1, 1, 6, 0\} \\
&B$_{10}$(2,2,$+ \sqrt{3 \over 2}$, 1) &\{ 1, 1, 6, 0\} \\
&B$_{11}$(1,1,0,8) &\{ 0, 0, 0, 3\} \\
&B$_{-}$(1,1,0,1) &\{ 0, 0, 0, 0\} \\
&B$_{+}$(1,1,0,1) &\{ 0, 0, 0, 0\}  \\ \hline \hline
\end{tabular}

\end{center}
}

\begin{center}
Table IIc-(2). \\
\begin{tabular}{ ccc } \hline\hline
 G$_{123}$ submultiplet  \\ \hline
$\phi$, R$_{1}$, L$_{1}$ in Table IIa-(2)  \\ \hline \hline
\end{tabular}

\end{center}

\begin{center}
Table IId-(1). \\
\begin{tabular}{ ccc } \hline\hline
SO(10) representation & G$_{2213}$ submultiplet &
\{$\lambda_{2L}^{U}$,$\lambda_{2R}^{U}$,$\lambda_{1X}^{U}$,$\lambda_{3C}^{U}$\}
\\
\hline
{\bf 10} &T$_{1}$, T$_{2}$ & \\
{\bf 126} & All H's in Table IIc-1 & \\
          &H$_{\Delta}$(3,1,$\sqrt{3 \over 2}$,1) &\{ 4, 0, 9, 0\} \\
{\bf 45} &S$_{1}$(1,1,0,8) &\{ 0, 0, 0, 3\} \\
          &S$_{2}$(3,1,0,1) &\{ 2, 0, 0, 0\} \\
          &S$_{3}$(1,3,0,1) &\{ 0, 2, 0, 0\} \\
          &S$_{-}$(1,1,0,1) &\{ 0, 0, 0, 0\}   \\ \hline \hline
\end{tabular}

\end{center}

\begin{center}
Table IId-(2). \\
\begin{tabular}{ ccc } \hline\hline
 G$_{123}$ submultiplet  \\ \hline
$\phi$, R$_{1}$ in Table IIa-(2) \\ \hline \hline
\end{tabular}

\end{center}

\begin{center}
Table III. \\
\begin{tabular}{ cccc } \hline\hline
Model & M$_{I}$ (GeV) & M$_{U}$ (GeV) & $\alpha_{U}^{-1}$ \\ \hline
A & 10$^{13.64}$  & 10$^{15.02 \pm .25}$  & 40.76 $\pm .16$ \\

B  & 10$^{10.70}$  & 10$^{16.26 \pm .25}$  & 46.35 $_{-.22}^{+.23}$  \\

C &10$^{10.16}$  & 10$^{15.55 \pm .20}$  & 43.86 $\pm .18$   \\
D &10$^{9.08}$  & 10$^{16.42 _{-.22}^{+.23}}$  & 46.12 $_{-.16}^{+.15}$  \\
\hline \hline
\end{tabular}

\end{center}
\newpage
\begin{center}
Table IV. \\
\begin{tabular}{ c||c|c|c|c } \hline\hline
Threshold Uncertainty & Model A  & Model B  & Model C & Model D \\ \hline
 M$_{I}$/M$_{I}^{0}$

   & 10$^{\pm 0.886}$

	   & 10$^{_{-0.067}^{+2.658}}$

	   & 10$^{\pm 0.571}$

	   & 10$^{_{-0.303}^{+0.690}}$  \\
 M$_{U}$/M$_{U}^{0}$ 	  & 10$^{\pm 0.481}$

          & 10$^{_{-1.240}^{+0.131}}$

          & 10$^{\pm 0.031}$

          & 10$^{_{-0.179}^{-0.138}}$ \\ \hline
 M$_{I}$/M$_{I}^{0}$  & 10$^{\pm 0.886}$

          & 10$^{_{-0.067}^{+2.658}}$

          & 10$^{\pm 0.000}$

          & 10$^{_{+0.083}^{+0.303}}$ \\
 M$_{U}$/M$_{U}^{0}$  & 10$^{\pm 0.481}$

          & 10$^{_{-1.240}^{+0.131}}$

          & 10$^{\pm 0.429}$

          & 10$^{_{-0.497}^{+0.179}}$    \\  \hline\hline
\end{tabular}

\end{center}


\begin{thebibliography}{10}

\bibitem{rf1}
H.~Georgi, in {\em Particles and Fields}, edited by C.E.~Carlson
 (American Institute of
Physics, New York, 1975);
H.~Fritzsch and P,~Minkowski, {\em Ann. Phys. (N.Y.) {\bf 93}, 193} (1975).

\bibitem{rf2}
N. ~Hata and P.~Langacker,
 {\em Univ. of Pensylvania Report UPR-0592T} (1993, to be published in
{\em Phys. Rev. {\bf D}}) ;
P.I.~Krastev and S.T.~Petcov, {\em Report No.SISSA 177/93/EP} (1993,
unpublished);  G. Fogli, E. Lisi and D. Montanino, CERN-TH.6944/93.

\bibitem{rf3}
R.~Davis et al.,{\em Proceedings of the 21st International Cosmic Ray
Conference}, Vol.~12, edited by R.J.~Protheroe (Univ. of Adelaide Press,
Adelaide, 1990), p.~143; K.S.~Hirata et al., {\em Phys. Rev. {\bf D 44}, 2241}
(1991);
A.I.~Abazov et al., {\em Phys. Rev. Lett. {\bf 67}, 3332} (1991) and
 T. Bowles, Invited talk at ICNAPP, Bangalore (1994);
 P.~Anselmann et al., {\em Phys.
Lett. {\bf B285}, 376} (1992) and {\em {\bf 314B}, 445} (1993).
T. Kirsten, Invited talk at ICNAPP, Bangalore (1994).

\bibitem{rf4}
For a theorectical discussion,
 see J.N.~Bahcall and M.H.~Pinsonneault, {\em Rev. Mod.
Phys. {\bf 64}, 885} (1992);
S.~Turck-chieze et al, {\em Phys. Rep. {\bf 230}, 57} (1993)

\bibitem{rf5}
S.P.~Mikheyev and A.Yu.~Smirnov, {\em Yad. Fiz., {\bf 42}, 1441}
(1975);
L.~Wolfenstein, {\em Phys. Rev. {\bf D 17}, 2369} (1978).

\bibitem{rf6}
K.~S. Babu and R.~N. Mohapatra,
\newblock {\em Phys. Rev. Lett. {\bf 70}, 2845} (1993).

\bibitem{rf7}
M.~Gell-Mann, P.~Ramond, and R.~Slansky,
\newblock {\em In Supergravity}, edited by D. Freedman et. al. (North-Holland,
  Amsterdam, 1980);
T.~Yanagida,
\newblock {\em Proceedings of the KEK workshop, 1979} (unpublished);
R.~N. Mohapatra and G.~Senjanovi\'{c},
\newblock {\em Phys. Rev. Lett. {\bf 44}, 912} (1980).

\bibitem{rf8}
M.~Fukugita and T.~Yanagida,
\newblock {\em Phys. Lett. {\bf B174}, 45} (1986);
P.~Langacker, R.D.~Peccei and T.~Yanagida,
\newblock {\em Mod. Phys. Lett. {\bf A1}, 541} (1986);
M.~Luty,
\newblock {\em Phys. Rev. {\bf D45}, 455} (1992);

\bibitem{rf9}
M.~Chanowitz, J.~Ellis and M.K.~Gaillard,
\newblock {\em Nucl. Phys. {\bf B135}, 66} (1978).

\bibitem{rf10}
D.~Chang, R.N.~Mohapatra, and M.K.~Parida,
\newblock {\em Phys. Rev. Lett. {\bf 52}, 1072} (1984).

\bibitem{rf11}
G.~Lazaridis, M.~Magg and Q.~Shafi,
\newblock {\em Phys. Lett. {\bf B97}, 87} (1980);
F.~Buccella, L.~Cocco and C.~Wetterich,
\newblock {\em Nucl. Phys. {\bf B248}, 273} (1984).

\bibitem{rf12}
D.~Chang and A.~Kumar,
\newblock {\em Phys. Rev. {\bf D33}, 2695} (1986).
J.~Basaq, S.~Meljanac, and L.~O'Raifeartaigh,
\newblock {\em Phys. Rev. {\bf D39}, 3110} (1989);
X.-G.~He and S.~Meljanac,
\newblock {\em Phys. Rev. {\bf D40}, 2098} (1989).

\bibitem{rf13}
\"{O}.~Kaymakcalan, L.~Michel, K.C.~Wali, W.D.~McGlinn, and L.~O'Raifeartaigh,
\newblock {\em Nucl. Phys. {\bf B267}, 203} (1986);
R.~Thornburg and W.D.~McGlinn,
\newblock {\em Phys. Rev. {\bf D33}, 2991} (1986);
R.~Kuchimanchi,
\newblock {\em Phys. Rev. {\bf D47}, 685} (1993).

\bibitem{rf14}
A.~Suzuki,
\newblock {\em KEK Preprint. 93-96} (1993).

\bibitem{rf15}
S.~Weinberg,
\newblock {\em Phys. Lett. {\bf B91}, 51} (1980);
L.~Hall,
\newblock {\em Nucl. Phys. {\bf B178}, 75} (1981).

\bibitem{rf16}
D.~Chang, R.N.~Mohapatra, J.M.~Gipson, R.E.~Marshak, and M.K.~Parida.
\newblock {\em Phys. Rev. {\bf D31}, 1718} (1985).

\bibitem{rf17}
N.G.~Deshpande, E.~Keith, and P.~B. Pal.
\newblock {\em Phys. Rev. {\bf D46}, 2261} (1992).

\bibitem{rf18}
F.~del Aguila and L.~Ib\'{a}$\tilde{n}$ez,
\newblock {\em Nucl. Phys. {\bf B177}, 60} (1981);
R.~N. Mohapatra and G.~Senjanovi\'{c},
\newblock {\em Phys. Rev. {\bf D 27}, 1601} (1983).

\bibitem{rf19}
F.~Buccella, G.~Miele, L.~Rosa, P.~Santorelli, and T. Tuzi,
\newblock {\em Phys. Lett. {\bf B233}, 178} (1989).

\bibitem{rf20}
D.-G.~Lee, (1993, Submitted to {\em Phys. Rev. {\bf D}}).

\bibitem{LEP}
The L3 collaboration, {\em Phys. Rep.} {\bf 236}, 1 (1993).

\bibitem{rf21}
R.~N. Mohapatra and M.~K. Parida,
\newblock {\em Phys. Rev. {\bf D47}, 264} (1993).

\bibitem{rf22}
M.~K. Parida and P.~K. Patra,
\newblock {\em Phys. Rev. Lett. {\bf 66}, 858} (1991) and {\em {\bf 68}, 754}
(1992).


\bibitem{rf23}
L.~Lavoura and L. Wolfenstein,
\newblock {\em Phys. Rev. {\bf D48}, 264} (1993).

\bibitem{rf24}
P.~Langacker,
\newblock {\em In Inner Space and Outer Space}, edited by E.~Kolb et. al.
(University of
Chicago Press, 1986), p.1.

\bibitem{rf26}
The predictions for cases B and D are different from those given in Ref.[22],
because of different input LEP data.

\end{thebibliography}
\end{document}